\documentstyle [aps,prl,twocolumn]{revtex}
\input{epsf}

\begin{document}
\twocolumn[\hsize\textwidth\columnwidth\hsize\csname@twocolumnfalse\endcsname
\title{The Universal Gaussian in Soliton Tails}
\author{David A. Kessler\cite{emailk}}
\address{Minerva Center and Department of Physics,
Bar--Ilan University, Ramat Gan 52900, Israel}
\author{Jeremy Schiff\cite{emails}}
\address{Department of Mathematics and Computer Science,
Bar--Ilan University, Ramat Gan 52900, Israel}
\maketitle
\begin{abstract}
We show that in a large class of equations, solitons formed from 
generic initial conditions do not have infinitely long exponential
tails, but are truncated by a region of Gaussian decay. This phenomenon
makes it possible to treat solitons as localized, individual objects.
For the case of the KdV equation, we show how the Gaussian decay emerges in the
inverse scattering formalism.
\end{abstract}
]

Recently, progress has been made understanding the time development
of the leading exponential edge in propagating fronts \cite{DAK}.
Inspired by results on fronts in the Fisher equation \cite{BD,Wim},
it was shown in \cite{DAK} that in generic reaction-diffusion equations like
Ginzburg-Landau (GL), an initial condition with a compact front gives rise to
a front with a leading exponential edge that does not extend forever,
but rather is cut off a finite distance ahead of the front.
The need for this is clear on physical grounds: the front, if one
is sufficiently far ahead of it, has not had time to make its presence felt,
and so the field exhibits the typical Gaussian falloff of the Green's
function (which lacks an intrinsic scale).  It was discovered that there
is a well-defined transition region
of width $O(\sqrt{t})$ wherein the field crosses over from the steady-state
exponential to Gaussian falloff.  This ``precursor'' transition region
propagates out ahead of the front, with a velocity $c^*$ which is greater
than twice the velocity $c$ of the front itself.

Given the basic underlying physics, the existence of such transition
regions must be a very general phenomenon, true not just of
reaction-diffusion fronts, but of other propagating solutions, such as 
the solitons in the KdV equation,
\begin{equation}
u_t=-u_{xxx}+6uu_x\ .
\end{equation}
The known exact one-soliton solutions
take the form $u(x,t)=-\frac12c\ {\rm sech}^2(\frac{1}{2}\sqrt{c}(x-x_0-ct))$,
and exhibit exponential decay for large $x$. The inverse scattering
transform \cite{DJ} tells us that the solution of KdV with generic initial
condition $u(x,0)=\phi(x)$ (with $\phi(x)\rightarrow 0$ sufficiently rapidly
as $\vert x \vert \rightarrow\infty$) consists of a train of solitons moving
to the right, along with a dispersive wave travelling to the left.
As above, we can argue that when $\phi(x)$  has compact support,
the solution emerges from the rightmost soliton decaying as
$\exp(-\sqrt{c}x)$, where $c$ is the relevant speed, but for
sufficiently large $x$ the presence of the solitons will not yet be felt,
and the behavior of the solution will be determined
by the Green's function of the linearized equation $u_t=-u_{xxx}$,
i.e. $u$ will decay roughly as $\exp(-2x^{3/2}/3\sqrt{3t})$ \cite{A}. 
Thus there is a transition in the nature of the decay.
How and where does this transition take place?

If there is more than one soliton in the soliton train, say two, with speeds
$c_1,c_2$ ($c_2>c_1>0$), then a further problem arises.
The solution emerges from the faster-moving soliton decaying as
$\exp(-\sqrt{c_2}x)$, but because the tail of the slower-moving soliton
falls slower, it is possible that it will return to
dominate, i.e. the decay will slow to $\exp(-\sqrt{c_1}x)$.
(For very large $x$, as explained above, the solution must go roughly as
$\exp(-2x^{3/2}/3\sqrt{3t})$.) We note that exact two-soliton solutions
\begin{equation}
 u(x,t) = -  \frac{(c_2-c_1)(c_2\cosh^2\alpha_1+c_1\sinh^2\alpha_2)}
    {2(\sqrt{c_2}\cosh\alpha_1\cosh\alpha_2-\sqrt{c_1}\sinh\alpha_1
                \sinh\alpha_2)^2}\ ,
\end{equation}
where $\alpha_1 = \frac{1}{2}\sqrt{c_1}(x - c_1 t)$ and
$ \alpha_2 = \frac{1}{2}\sqrt{c_2}(x - c_2 t)$,
exhibit exactly this phenomenon: when $x$ is large $u\sim\exp(-\sqrt{c_1}x)$,
i.e. the tail of the slow soliton dominates the decay. This is clearly
undesirable in physical situations, as it means solitons cannot be considered
as isolated objects.

In this paper we study the tails of generic solitons, i.e. those produced
by taking a generic compactly supported (or very rapidly decaying)
initial condition for KdV. The results are quite
striking. If the soliton is centered at $x=ct$, then in a region
of width $O(c^{1/4}t^{1/2})$ around $x=c^*t$, where $c^*=3c$,
there is a rapid transition in
the behavior of the tail, and the decay changes from exponential to
Gaussian. This behavior
persists until $x-c^*t=O(c^{1/2}t^{2/3})$, and then there is a second
transition to the final region, in which $u$ decays roughly as
$\exp(-2x^{3/2}/3\sqrt{3t})$. In the Gaussian region
the decay is very rapid, and the soliton tail is effectively cut off,
making the soliton an isolated object.

This behavior is predicted by analysis of the linearized equation
$u_t=-u_{xxx}$ alone, and does not involve any of the special properties
of the KdV equation. It is thus {\em universal} for soliton solutions
of PDEs with this linearization, and the existence of
a Gaussian cut-off region is in fact universal in a much larger class
of equations, and is responsible, as we have explained, for the individuality
of solitons. The existence of exact two-soliton solutions in which
the solitons are not genuinely separate objects is one of the special
properties of the KdV equation, and is not physical.


The rest of this paper proceeds as follows: We first give the
arguments for the behavior of the tail outlined above. We then
show how this makes the soliton an isolated object. In the last
part, we connect our results to the exact Inverse Scattering solution,
and use this to generate a numerical solution of the KdV equation,
confirming the picture presented.



{\em Soliton Tails ---}
Since we only want to look at the soliton tail, where $u$ is small,
we work with the linearized equation $u_t=-u_{xxx}$. We have
explained in the introduction why the standard soliton tail
$u=-2c\exp(-\sqrt{c}(x-ct))$ is not a physically acceptable solution to
this; we require a solution that for large $x$ decays faster.
We look for an acceptable  solution in the form
\begin{equation}
u(x,t) = -2c\exp(-\sqrt{c}(x-ct)) f(y,t)
\label{form}
\end{equation}
where $y\equiv x-c^* t$ and $c^*$ is a constant to be determined, and where
$f$ satisfies
boundary conditions $f\rightarrow 1$ as $y\rightarrow -\infty$ and
$f\rightarrow 0$ as $y\rightarrow +\infty$.
Substituting in, we find $f$ satisfies
\begin{equation}
f_t = -f_{yyy} + 3\sqrt{c} f_{yy} + (c^*-3c) f_y\ .
\label{basic}
\end{equation}
We choose $c^*=3c$ so that the term in $f_y$ drops out. The key point is
that at large times the diffusive term, $3\sqrt{c}f_{yy}$, dominates the RHS
out to $y$ of order $c^{1/2}t^{2/3}$, whereas the dispersive term, $f_{yyy}$,
dominates for $y\gg ct$. To see this, let us first drop the $f_{yyy}$ term.  
The resulting diffusion equation then has the exact scaling solution
\begin{equation}
f(y,t)=\frac{1}{2}\text{erfc}\left(\frac{y}{2\sqrt{3}c^{1/4}t^{1/2}}\right)
\label{erfc}
\end{equation}
which can be seen to satisfy the boundary conditions.  This solution
transforms
the original exponential falloff of $u$ to a Gaussian falloff in a region
of width $O(c^{1/4}t^{1/2})$ around $y=0$, or equivalently $x=3ct$.  This
{\em erfc} cutoff of the exponential moving out ahead of the front is
exactly what occurs in the case of the GL equation \cite{DAK}; there
it is an exact solution of the relevant linearized equation.

Here, however, we have to address the effect of the neglected $f_{yyy}$ term.
There are two cases, depending on the size of the scaling variable
$z\equiv y/c^{1/4}t^{1/2}$. For $z=O(1)$, the $f_{yyy}$ term is
down by a factor $c^{-3/4}t^{-1/2}$ and so is in fact negligible for large
$t\gg c^{-3/2}$. For large $z$, on the other hand, the
$f_{yyy}$ term induces a correction of order
$z^3/c^{3/4}t^{1/2}$ and so once $z=O(c^{1/4}t^{1/6})$,
or equivalently $y=O(c^{1/2}t^{2/3})$, it is no longer negligible,
as we noted above.
In fact, for very large $y\gg ct$, the $f_{yyy}$ dominates,
and the controlling
factor in $f$ is the $\exp(-y^{3/2}/t^{1/2})$ of the Green's function,
as expected.  In this regime, the diffusive term can be seen to be of
subleading order, inducing a correction
to the argument of the exponential of order $\sqrt{c}y$.

The upshot of this is that the cutoff of the exponential is
provided by the {\em erfc}, inducing Gaussian decay.  Only much later does
the Gaussian decay slow down to that prescribed by  the Green's function.
It is clear from the very general nature of this argument that this
scenario is {\em universal}, applying to a wide range of soliton
equations, as well as front propagation problems like GL.

We have shown that the {\em erfc} solution is consistent, but because the
basic equation (\ref{basic}) is linear we can go further and prove that it
in fact is what arises from solving the initial value problem for
compact initial conditions.  To do this, we solve (\ref{basic})
by taking a Fourier transform in space, finding
\begin{equation}
f(y,t) = \frac{1}{2\pi}\int_{-\infty+i\epsilon}^{\infty+i\epsilon}
    e^{iky-(3\sqrt{c}k^2-ik^3)t} \tilde{f}(k) dk\ , \label{sol}\end{equation}
where $\tilde{f}(k)$ is the Fourier transform of the initial
condition:
\begin{equation}
\tilde{f}(k) = \int_{-\infty}^{\infty} f(y,0)e^{-iky} dy\ .
\end{equation}
It will turn out that in the region of $y \ll  ct$, the only
thing we need to know about $f$ is its behavior at small $k$.
Since the long-distance structure of $f(y,0)$ is that of a
step-function $\theta(-y)$, the small $k$ behavior of $\tilde f$
is precisely that of the Fourier transform of $\theta(-y)$, i.e.
for small $\vert k\vert$
\begin{equation}
\tilde{f}(k) \approx \frac{i}{k}\ , \qquad  {\rm Im}(k)>0\ .
\label{fapprox}\end{equation}
The range of validity of this approximation is $\vert k \vert\ll a^{-1}$,
where $a$ is a characteristic length scale of $f(y,0)$.

We now use saddle point techniques to evaluate the integral (\ref{sol}).
We denote the factor in the exponential in the integrand by $g(k)$, i.e.
$g(k)=iky-(3\sqrt{c}k^2-ik^3)t$. For all
positive $t$ and $y>-3ct$, $g(k)$ has  two critical points on the imaginary 
axis, at $k_{\pm}=i(\pm\sqrt{c+y/3t}-\sqrt{c})$.  For $y>0$ we deform the 
integral in (\ref{sol}) to the steepest descent integral coming in along
the ray ${\rm arg}(k)=5\pi/6$, going through the critical point
on the positive imaginary axis at $k=k_+$, and going out along the ray
${\rm arg}(k)=\pi/6$ (along both these rays $ik^3$ is negative
real). Writing $\lambda=(3xt)^{1/4}(k-k_+)$, the factor in the exponential
is $g(k)=g(k_+)-\lambda^2+i\lambda^3(t/27x^3)^{1/4}$; provided $t/x^3$
is small, which it will be, for example, for positive $y$ and 
$t\gg c^{-3/2}$, we can ignore the term in $\lambda^3$. Turning now to
the factor multiplying the exponential in (\ref{sol}), this is 
$\tilde{f}(k)=\tilde{f}(k_++(3xt)^{-1/4}\lambda)$. Provided $\vert k_+\vert$ 
and $(3xt)^{-1/4}$ are sufficiently small
($y\ll \min(ct,\sqrt{c}a^{-1}t)$ and $t\gg a^2c^{-1/2}$ are sufficient 
conditions), we can use the approximation (\ref{fapprox}) to estimate this. 
Putting all the approximations together, we have that for $y>0$,
$y\ll \min(ct,\sqrt{c}a^{-1}t)$ and $t\gg \max(a^2c^{-1/2},c^{-3/2})$:
\begin{eqnarray}
f(y,t)&\approx&\frac{1}{2\pi}e^{g(k_+)}
  \int_{-\infty}^{\infty} \frac{ie^{-\lambda^2}}
   {\lambda + (3xt)^{1/4}k_+} d\lambda\ \nonumber \\
  &=& \frac12 \exp\left(g(k_+)+(3xt)^{1/2}\vert k_+\vert^2 \right) 
     \nonumber\\
 && ~~~   {\rm erfc}\left((3xt)^{1/4}\vert k_+\vert \right)\ .
\label{experfc}\end{eqnarray}
(See for example \cite{GR} 3.466 for the necessary integral.) 
This expression can be substantially simplified provided
$y\ll 6c^{1/2}t^{2/3}$, in which case the factor in
the exponential in (\ref{experfc}) is $o(1)$, and  we recover (\ref{erfc}). 
A similar calculation recovers (\ref{erfc}) in the case $y<0$ as well.

We can also use the Fourier integral (\ref{sol}) to learn more about the
post-Gaussian region.  As $y$ gets larger, so does $\vert k_+\vert $ and we
can no longer use the small-$k$ limit of $\tilde f$.  The physical
meaning of this is that this region is sensitive to the exact
initial conditions. This must be so, since the large $y$ regions feel
only the initial condition, since no other information has had
time to propagate there.  Nevertheless, we can obtain the controlling
factors in $f(y,t)$.  Once we are sufficiently far
from the $k=0$ pole in $\tilde f$, we can treat $\tilde f$ as a constant
when doing the saddle point integral.  Using the exact form of $k_+$
above we obtain
\begin{equation}
f(y,t) \approx \frac{\tilde{f}(k_+)}{2\pi^{1/2}(3tx)^{1/4}}
   \exp\left(-\frac{2x}{3}\sqrt{\frac{x}{3t}}+\sqrt{c}(x-ct)\right)
\label{largey}
\end{equation}
This equation is valid for in the regime $t\gg \max(a^2c^{-1/2},
\linebreak[0]c^{-3/2})$ 
and $z\gg 1$, and also in the regime $y\gg \max(a^4t^{-1},
\linebreak[0]c^{1/4}t^{1/2})$, 
$t$ arbitrary. 

We see that the two approximations, (\ref{erfc}) and (\ref{largey}) have
a region of overlap for large $t$, namely $c^{1/4}t^{1/2} \ll  y \ll
c^{1/2}t^{2/3}$.  This is precisely the Gaussian region. Either by
using $y\ll ct$ and the pole approximation for $\tilde{f}$ in 
(\ref{largey}) or by using the large argument approximation of {\em erfc}
in (\ref{erfc}) we recover
\begin{equation}
f(y,t) \sim \frac{c^{1/4}(3t)^{1/2}}{\sqrt{\pi}y}
     \exp\left(-\frac{y^2}{12\sqrt{c}t}\right)\ .\label{gaussian}
\end{equation}

{\em Soliton Individuality ---}
As explained in the introduction, if solitons had infinitely long
exponential tails, then in a situation where there were two solitons,
with speeds $c_1,c_2$ ($c_2>c_1>0$), the tail of the slow soliton would
dominate the large $x$ decay. In this section we show that the cutting off
the tail of a soliton with speed $c$ at $x=3ct$ (up to an additive constant)
prevents this.

Assuming exponential tails, the point $x(t)$ where the slow soliton
tail returns to dominate is determined by an equation of the form
$\sqrt{c_1}(x(t)-c_1t) = \sqrt{c_2}(x(t)-c_2t) + {\text{\em cnst}}$,
and so
\begin{equation}
\frac{dx(t)}{dt} = \frac{c_1^{3/2}-c_2^{3/2}}{c_1^{1/2}-c_2^{1/2}}
  = c_1+\sqrt{c_1c_2}+c_2\ .
\end{equation}
Thus the putative point of ``return of the slow soliton'' must
move with a speed $c_1+\sqrt{c_1c_2}+c_2$.
But this exceeds $c_1^*=3c_1$, so for sufficiently large time the
slow soliton has been cut off before it can return to dominate the decay.

The meaning of this, as emphasized in the introduction, is that solitons
can genuinely be regarded as isolated objects; this is only due to the
Gaussian cut-off behavior we have discussed.

{\em Connection to Inverse Scattering ---}
The KdV equation can be solved via the Inverse
Scatting Transform (IST).  This should reproduce
the results obtained by our general arguments above. 
For large $x$, where the field $u(x,t)$ is
small, the IST tells us that 
$u\approx 2\frac{\partial}{\partial x}F(x,t)$, where
\begin{equation}
F(x,t) = \sum_{n=1}^Nc_n^2e^{8\kappa_n^3t-2\kappa_nx} +
   \frac{1}{2\pi}\int_{-\infty}^{\infty} b(k)e^{8ik^3t+2ikx} dk\ .
\label{F1}\end{equation}
The various constants in this equation are scattering data
for the Schr\"odinger operator
$-\partial_x^2+\phi(x)$ associated with the initial condition $\phi(x)$;
specifically $-\kappa_n^2$ gives the discrete spectrum
($n=1,\ldots,N$), the $c_n$  are the normalization constants
and $b(k)$ is the reflection coefficient.

When $\phi(x)$ is a reflectionless potential, $b(k)$ vanishes and
exact soliton solutions are obtained.  However, for generic initial conditions,
including the compact initial conditions considered here, $b(k)$ does
not vanish.  In fact, it has a number of poles on the positive imaginary
axis, which are associated with the  solitons that develop.
When we move the integration path in (\ref{F1}) to a new contour ${\cal C}$
above all these poles,
the residues exactly cancel the discrete sum in (\ref{F1}) \cite{DJ},
leaving, after a rescaling of $k$ by a factor 2,
\begin{equation}
u(x,t)= \frac{i}{2\pi}\int_{\cal C} \exp(\ln b(k/2)+ik^3t+ikx) k dk \ ,
\label{ist}
\end{equation}
Translating $k$ by $2i \kappa_1$, where $i\kappa_1$ is the location of the
uppermost pole of $b(k)$ (which gives rise to the fastest soliton, with 
velocity $c=4\kappa_1^2$), we obtain the solution in exactly the 
form given by equations (\ref{form}) and (\ref{sol}), where $\tilde f(k)$
is identified with $(\kappa_1-ik/2)b(k/2+i\kappa_1)/c$. As expected, this
$\tilde f(k)$ has a pole at zero, and no other singularities
in the upper-half $k$-plane.  The only trivial
difference is that the residue at the pole at zero
is not $i$, but some multiple thereof, which
lets us calculate the phase shift of the soliton, 
information not accessible in the general framework.

We can actually use (\ref{ist}) to numerically calculate $u$ for
a given initial condition, and thus verify our analytic arguments.
We present results for the case where $\phi(x)$ is a square well
of depth $V$ and
width $2a$ centered around $x=0$; for this case we have
\begin{equation}
 b(k) = \frac{-Ve^{-2iak}}
   {(q+ik\cot qa)(q-ik\tan qa)}\ ,\qquad  q = \sqrt{V+k^2}
   \ .
\end{equation}
The number of bound states (solitons)
is $1+[2\sqrt{V}a/\pi]$, associated with poles on the
imaginary axis of $b(k)$. The poles all have $\vert k
\vert < \sqrt{V}$.

\begin{figure}
\centerline{\epsfxsize=3.25in \epsffile{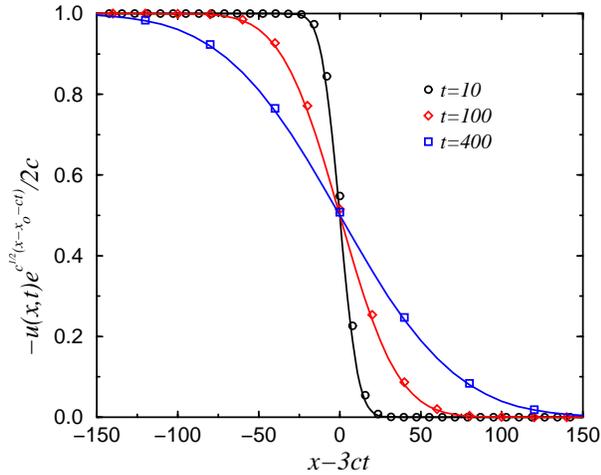}}
\caption{$f(y,t)=-u(x,t)\exp(c^{1/2}(x-x_0-ct))/2c$ vs. 
$y=x-3ct$ for $t=10$ (circle), 100 (diamond) and 400 (square), starting
from the square-well initial condition, $u(x,0)=-V\theta(a-|x|)$, with
$V=a=1$. The solid line is the analytic approximation, (\ref{erfc}).}
\end{figure}

Separating the exponent in (\ref{ist}) into its real and imaginary
parts, ${\cal A}(k) + i \Phi(k)$,
the idea is to integrate along the contour of constant-phase $\Phi=0$
passing through the saddle point which lies on the imaginary axis.
The integrand is symmetric about the imaginary axis, and is strictly
decreasing as we move away from the saddle.  Denoting our steepest-descent
contour by $k(s)\equiv \tau(s)+i\omega(s)$, parametrized by the arclength
$s$,
$u$ is given by $u(x,t)= 4 {\cal I}(\infty)/\pi $, where
\begin{equation}
{\cal I}(s)\equiv \int_0^s
ds' \left( \omega \frac{d\tau}{ds}  + \tau\frac{ d\omega}{ds}\right)
\exp({\cal A}(s'))\ .
\end{equation}
We simultaneously find
this steepest-descent contour and the integral $I(s)$ along the
contour by solving via
Runge-Kutta the following third-order system:
\begin{eqnarray}
\dot \tau &=& -\Phi_\omega/\sqrt{\Phi_\tau^2+\Phi_\omega^2}
               - \lambda \Phi \Phi_\tau/(\Phi_\tau^2+\Phi_\omega^2)
\nonumber\\
\dot \omega &=& +\Phi_\tau/\sqrt{\Phi_\tau^2+\Phi_\omega^2}
           - \lambda \Phi \Phi_\omega/(\Phi_\tau^2+\Phi_\omega^2)  \\
\dot {\cal I} &=& (-\Phi_\omega \omega + \Phi_\tau \tau) \exp({\cal A}) /
                         \sqrt{\Phi_\tau^2+\Phi_\omega^2} .\nonumber
\end{eqnarray}
Here the dot denotes a derivative with respect to $s$, and $\lambda$ is an
arbitrary (positive) Lagrange multiplier parameter stabilizing the $\Phi=0$
constraint.  It is easy to verify that $\dot \Phi(k(s)) = - \lambda \Phi$.
In practice,
$\dot {\cal I}$ decays rapidly in $s$, and the integration may be halted
after sufficient accuracy is achieved.  This process is easily repeated for
various $x$, $t$, yielding the results in Figs. 1 and 2.  In Fig. 1, we
plot $f(y,t)=-u(x,t)/2c\exp(-\sqrt{c}(x-x_0-ct))$ for $t=10,100,400$.  
We also exhibit our universal approximation for $f$, (\ref{erfc}).  
The argument is good for the smaller $t$, and excellent for the larger $t$'s.
 In Fig. 2, we again plot $f(y,t)$, this time in semi-log scale, along with 
our approximations (\ref{erfc}) and (\ref{largey}).  We see that the 
{\em erfc} works some way past the exponential cutoff, and that 
(\ref{largey}) works from the Gaussian regime outward, in accord with our
analytical understanding.

In summary, we have found that in the KdV equation, and many
other systems possessing solitons, the leading exponential edge of the
soliton is only built up over time.  At any given time, it is cut off at
some point by
a multiplicative {\em erfc} factor, which transforms the exponential decay
into a Gaussian falloff. The cutoff point moves out with time, so that the
length of exponential edge increases linearly with time.  This cutoff 
phenomenon serves to give the soliton an individual identity.

\begin{figure}
\centerline{\epsfxsize=3.25in \epsffile{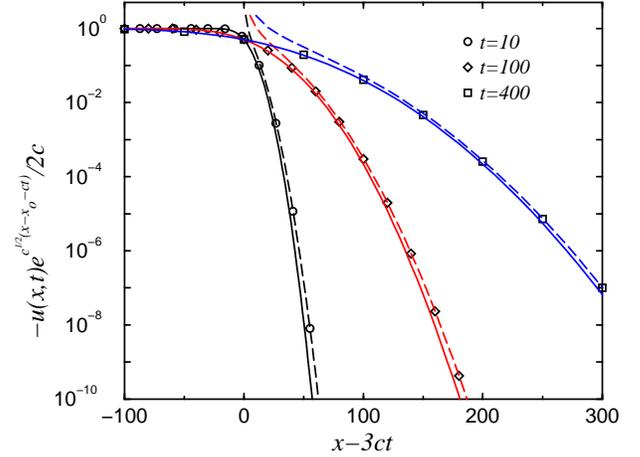}}
\caption{Same data as in Fig. 1, in semi-log scale, plotted together with 
the two overlapping approximations, (\ref{erfc}) (solid line) and 
(\ref{largey})
(dotted line).}
\end{figure}

\end{document}